# Determination of the 3γ fraction from positron annihilation in mesoporous materials for symmetry violation experiment with J-PET scanner.

B. Jasińska[1*], M. Gorgol[1], M. Wiertel[1], R. Zaleski[1], D. Alfs[2], T. Bednarski[2], P. Białas[2], E. Czerwiński[2], K. Dulski[2], A. Gajos[2], B. Głowacz[2], D. Kamińska[2], Ł. Kapłon[2,3], G. Korcyl[2], P. Kowalski[4], T. Kozik[2], W. Krzemień[5], E. Kubicz[2], M. Mohammed[2], Sz. Niedźwiecki[2], M. Pałka[2], L. Raczyński[4], Z. Rudy[2], O. Rundel[2], N.G. Sharma[2], M. Silarski[2], A. Słomski[2], A. Strzelecki[2], A. Wieczorek[2,3], W. Wiślicki[4], B. Zgardzińska[1], M. Zieliński[2], P. Moskal[2]

[1] *Department of Nuclear Methods, Institute of Physics, Maria Curie-Sklodowska University, Pl. M. Curie-Skłodowskiej 1, 20-031 Lublin, Poland*

[2] *Faculty of Physics, Astronomy and Applied Computer Science, Jagiellonian University, S.Łojasiewicza 11, 30-348 Kraków, Poland*

[3] *Institute of Metallurgy and Materials Science of Polish Academy of Sciences, W. Reymonta 25, 30-059 Kraków, Poland*

[4] *Świerk Computing Centre, National Centre for Nuclear Research, A. Soltana 7, 05-400 Otwock-Świerk, Poland*

[5] *High Energy Physics Division, National Centre for Nuclear Research, A. Soltana 7, 05-400 Otwock-Świerk, Poland*

**Abstract**

Various mesoporous materials were investigated to choose the best material for experiments requiring high yield of long-lived positronium. We found that the fraction of 3γ annihilation determined using γ–ray energy spectra and positron annihilation lifetime spectra (PAL) changed from 20% to 25%. The 3gamma fraction and o-Ps formation probability in the polymer XAD-4 is found to be the largest. Elemental analysis performed using scanning electron microscop (SEM) equipped with energy-dispersive X-ray spectroscop EDS show high purity of the investigated materials.

*Corresponding author: bozena.jasinska@umcs.pl

**Introduction**

The Jagiellonian Positron Emission Tomography scanner (J-PET) has been recently built at the Jagiellonian University [1-9] using a novel approach proposed in references [10, 11]. The J-PET detector consists of 192 plastic scintillator strips arranged axially in three cylindrical layers with the cylinder inner diameter of 85 cm. The detector is equipped with a solely digital readout electronics [12, 13] and dedicated reconstruction software [2, 4, 6, 7, 8]. The detector opens perspectives for experimentation in the fields of nuclear medical imaging, physics and nanobiology. The physics research program will include tests of quantum mechanics and discrete symmetries in the decays of positronium atoms. The measurement of angular correlations between the positronium spin and momentum and the polarization vectors of photons emitted in its decay will allow investigations of symmetries such as time reversal (T), charge conjugation (C) and their combinations (CP and CPT). The latter (CP and CPT) were so far tested in the decays of positronium with the precision of about 0.2% [14,15]. The more precise description of planned experiments one can find in the paper by Moskal et al., this volume.

Positronium (Ps) is an exotic atom composed of two light particles: electron and positron. Positrons are created, among others, during β decay of radioactive nuclei. When passing through the matter, initially they lose their energy and as a thermalized particles can annihilate directly with electrons of the medium or create bound state – positronium. Ps can exist in two substates: para-positronium (p-Ps) and ortho-positronium (o-Ps) with anitiparallel and parallel spin orientation, respectively. Both states are unstable and decay with the mean lifetime values of 125 ps (p-Ps) and 142 ns (o-Ps). In the medium, however, o-Ps lifetime can be shortened even below 1 ns due to the possibility of positronium being trapped in the regions of lower electron density than the bulk material and higher than the vacuum. In these voids, o-Ps can annihilate not only by intrinsic decay but also by so called pick-off process [16] with one of the electrons with antiparallel spin from surrounding bulk material. Shortening of the o-Ps lifetime value is determined by the local electron density or, one can say, by the size of the void in which positronium is trapped. Assuming the void as a potential well one can correlate the volume size with the mean lifetime value of o-Ps [17-20]. Free positron annihilation occurs predominantly *via* decay into two gamma quanta, and only about 1/372 of direct positron annihilations results in the emission of the three gamma quanta. Due to the conservation laws, p-Ps decays into even number of gamma quanta, mainly 2γ, both in a vacuum and in a matter. In turn, atoms of o-Ps annihilate in the vacuum emitting mainly 3γ, while in the matter, ortho-positronium can, in addition, annihilate into 2γ quanta via the above mentioned pick-off process with one of the electrons of surrounding material with antiparallel spin orientation. In consequence, a fraction of o-Ps in the examined medium decays by intrinsic decay into 3 γ, while another one - into 2 γ *via* pick-off process. Fractional rate of these competing processes depends on the void size, the larger is void, the higher is the lifetime value and the higher is the 3γ fraction ($f_{3\gamma}$). Both gamma-rays from 2γ annihilation have the same energy 511 keV while gamma rays produced in 3γ decay have a continuous energy spectrum between 0 and 511 keV; the sum of the energies of these three photons is 1022 keV.

The precise method of the $f_{3\gamma}$ determination with monoenergetic positron beam was proposed by Mills [21]. This method requires two materials for callibration: one where Ps is not formed and the other in which close to100% of Ps formation is observed [21, 22]. It was applied many times for the determination of 3γ fraction in various porous materials [see for example 23-27].

In the J-PET experiments, positronium will be created by irradiation of porous materials with positrons emitted from radioactive isotopes as e.g. $^{22}$Na or $^{68}$Ge. To make the experiment as efficient as possible and to lower its systematic uncertainties it is crucial to use materials with high positronium production probability and long survival time of positronium inside the pores. The smaller probability of Ps annihilation in pick-off the closer is its lifetime to the one in the vacuum value. In previous studies described e.g. in articles [25,26] it was concluded that one of the best materials for the production of positronium are silica aerogels composed of silica strands with tens of nm pores in which empty space constitutes more than 90% of the total volume [28].

In this paper we present investigations of some selected group of porous substances to determine the best material which can be applied as a target for production of positronium in the experiments planned with the J-PET detector.

**Experimental**

Four samples of commercial aerogels (IC 3100, IC3110, IC3120 and LA1000) available from Cabot Corporation (NYSE: CBT), commercially available Amberlite XAD4 (CAS 37380-42-0) supplied by ROHM & HAAS (now Dow Chemical Co.) and silica porous material SBA-15 synthesised in the Faculty of Chemistry of Maria Curia Sklodowska University [29] were investigated in order to find the one in which $3\gamma$ fraction of annihilating positrons is the highest.

Samples were examined using PALS (Positron Annihilation Lifetime Spectroscopy) and gamma-ray spectroscopy. PAL spectra were collected using fast – slow delayed coincidence spectrometer equipped with two $BaF_2$ scintillation detectors. A time range of time-to-amplitude converter equals to 2 µs enabled measurements of o-Ps lifetimes in a whole range of possible values. Total count number per each spectrum was set to about $25 \times 10^6$.

The 5 µC $^{22}$Na positron source enclosed in Kapton envelope was placed between two layers of sample inside the vacuum chamber under the pressure of about $10^{-5}$ Pa. Each sample was initially degassed for 24 hours. All measurements were performed at room temperature. Spectra were analyzed using LT 9.2 program [30].

The γ ray spectra were detected with a high purity germanium detector (HPGe, relative efficiency 29.8%, FWHM 1.71 keV at 1332 keV) in a large energies range. Samples were placed in the same vacuum chamber like in PALS measurements.

Additionally, the sample compositions were examined using scanning electron microscope (SEM) equipped with the energy-dispersive X-ray spectroscopy (EDS) enabling determination of the material elemental composision. Chemical composition of the material is crucial in the experiments involving positronium in the magnetic field as the possible ortho-para conversion in the presence of some ions can be observed [31].

**Results and discussion**

During PAL spectra processing fitting of five components was required: the shortest lived ~125 ps is ascribed to p-Ps decay, the next one comes from positron free annihilation and three the longest lived with a lifetime above 1 ns – from o-Ps decay (Table 1). A fraction of o-Ps atoms annihilating with the emission of 3γ quanta may be expressed as:

$$f_{3\gamma}^{o-Ps} = \frac{\tau_{o-Ps}}{\tau_T},$$

where $\tau_{o-Ps}$ denotes the lifetime value of o-Ps in a respective sample and $\tau_T$ = 142 ns is o-Ps lifetime value in a vacuum. In the case when only a few kinds of free volumes/pores exist in the material then few components in PALS measurements come from o-Ps decay, the 3γ fraction in the whole spectrum can be effectively described by dependence:

$$f_{3\gamma} = \frac{\left(1-\sum_i P_i\right)}{372} + \frac{3}{4}\sum_i \frac{\tau_i(o-Ps)}{\tau_T}P_i$$

where $i$ denote the i-th o-Ps component, and $P_i$ denotes the probability of the i-th component formation. Positronium formation probability P can be in the simplest form calculated from o-Ps intensity as $P_i = 4/3 I_i$. Results of $f_{3\gamma}$ calculation are presented in Table 2. However, the method of 3γ fraction determination presented above can be only treated as a rough approximation, as in practice, the extracted intensities of o-Ps components depend on the efficiencies of detection of 2γ and 3γ events, which ratio changes with the width of the "stop" window of the PAL spectrometer.

Commonly accepted method of 3g fraction determination is the one based on the annihilation γ–ray spectra processing. In Fig. 1 spectra of a few selected porous materials are presented in comparison to the material in which Ps is not formed. As a reference material in this paper we used Kapton, where lifetime of 374 ps confirms no positronium formation. As we have no possibility to perform measurements in the temperature close to 1000K, we are not able to add spectrum for 100% of Ps formation in calculation (when using so called valley to 511-keV-peak method). Spectra were normalized to the same height of the 511 keV line in order to emphasize the differences in low energy region. In this analysis the $f_{3\gamma}$ was determined directly from reduction of the 511 keV peak in comparison to the reference material (0% Ps formation). Applying this method, spectra have to be normalized to the same number of β decays e.g. to the same area under 1274 keV peak which is proportional to the number of emitted positrons ($^{22}$Na → $^{22}$Ne* e$^+$ ν$_e$ → $^{22}$Ne γ(1274) e$^+$ ν$_e$). Then $f_{3\gamma}$ can be calculated basing only on the counts number under the 511 keV line. The difference in the background between the left and right hand side of peak, being the result of the Compton scattering effect, was also corrected. This method seems to be much more accurate in the case of measurements related to one reference (0% of Ps) only, as one can neglect the different detector efficiency of low and high energy quanta. Additionally, at high yield of 3γ fraction, the valley-to-511-peak ratio is not a linear function of $f_{3\gamma}$. From Table 2 comes that in our case $f_{3\gamma}$ calculated from PAL spectra is about 20% overestimated in comparison to $\gamma$ spectra. The highest fraction of 3γ events, close to 25% was found in two samples: an aerogel LA-1000 and a porous polymer XAD-4. It was presented elsewhere [32] that total yield of positronium was achieved up to 100% and the highest 3γ fraction was found being equal almost 50% [26] while in our case it is much lower. However, those measurements were performed for Ps outdiffusing from the sample to the vacuum (when the fraction of annihilating free positrons is negligible). In our case, Ps annihilated inside the sample. In this case two processes leading to 3γ annihilation take place, as it was presented in Introduction. For the studies of the symmetry violation in the o-Ps decays it is important that the contribution of 3γ events from o-Ps decay, $x=f_{3\gamma}(o\text{-}Ps)/f_{3\gamma}(e^+e^-)$, should be as high as possible. In polymer XAD-4 and aerogel LA-1000 x has the largest values as both, the o-Ps lifetime value and Ps formation probability are the highest in these materials.

For two samples mentioned above, the SEM images and elemental microanalysis were additionally performed (Fig. 2). Microanalysis exhibits that materials do not contain impurities even at the level of sensitivity of EDS (search of some elements like Fe was additionally included): aerogel LA-1000 is composed of silicon and oxygen (presence of carbon is connected to groundsheet used ), while polymer of carbon, oxygen, small amount of silicon (probably from aerogels pulverised in vacuum chamber); hydrogen is not detected by this technique.

**Conclusions**

Examined samples of porous materials (except IC3100) exhibit high fraction of positrons annihilating with emission of 3γ rays, from 20-25%. The highest fraction was detected in aerogel LA-1000 and polymer XAD-4 which are suitable for a target material to perform planned experiment concerning a symmetry violation in leptonic system. The results of SEM elemental microanalysis confirm this choice, because both materials do not contain impurities which can influence the Ps formation yield or ortho-para spin conversion in the presence of magnetic field.

Table 1. Lifetimes and intensities of o-Ps obtained from the Positron Annihilation Lifetime Spectra

| Sample | $I_3$ [%] | $\tau_3$ [ns] | $I_4$ [%] | $\tau_4$ [ns] | $I_5$ [%] | $\tau_5$ [ns] |
|---|---|---|---|---|---|---|
| IC3100 | 6.9(8) | 0.62(12) | 0.5(2) | 9.4(5) | 17.9(5) | 131.9(3) |
| IC3110 | 4.1(8) | 0.77(18) | 0.6(2) | 19.7(9) | 26.3(5) | 126.6(3) |
| IC3120 | 4.0(8) | 0.78(18) | 0.6(2) | 18.8(9) | 26.8(5) | 126.6(3) |
| LA1000 | 2.0(8) | 0.88(23) | 0.6(2) | 17.9(9) | 29.1(5) | 125.0(3) |
| SBA-15 | 1.1(6) | 2.52(39) | 2.5(6) | 54.8(2.5) | 27.4(6) | 126.4(9) |
| XAD-4 | 3.3(6) | 2.45(25) | 2.8(5) | 10.2(6) | 44.8(4) | 90.8(12) |

Table 2. The 3γ fraction determined from PALS and γ–ray spectra.

| Sample | $f_{3\gamma}$ [%], PALS | $f_{3\gamma}$ [%], γ–ray |
|---|---|---|
| IC3100 | 16.9(66) | 13.34(13) |
| IC3110 | 23.7(118) | 19.49(19) |

| IC3120 | 24.2(109) | 18.96(18) |
| LA1000 | 25.8(158) | 21.59(21) |
| SBA-15 | 25.5(173) | 19.80(19) |
| XAD-4 | 28.9(54) | 24.44(24) |

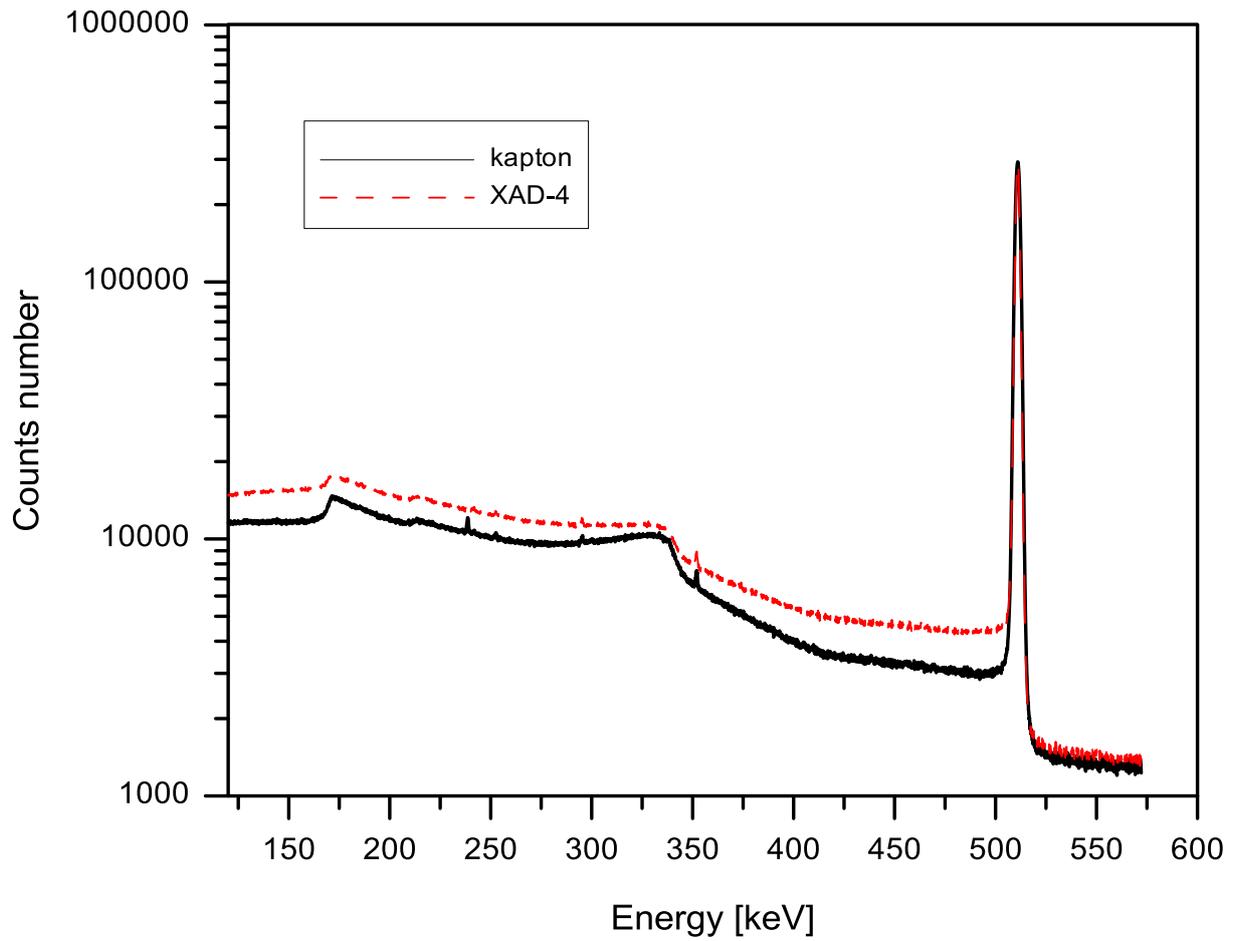

Fig. 1. Gamma–ray spectra measured using a HPGe detector. Kapton is reference material where no positronium is formed (0% Ps). Spectra were normalized to an equal height of the 511 keV peak. Small peaks come from natural radiation.

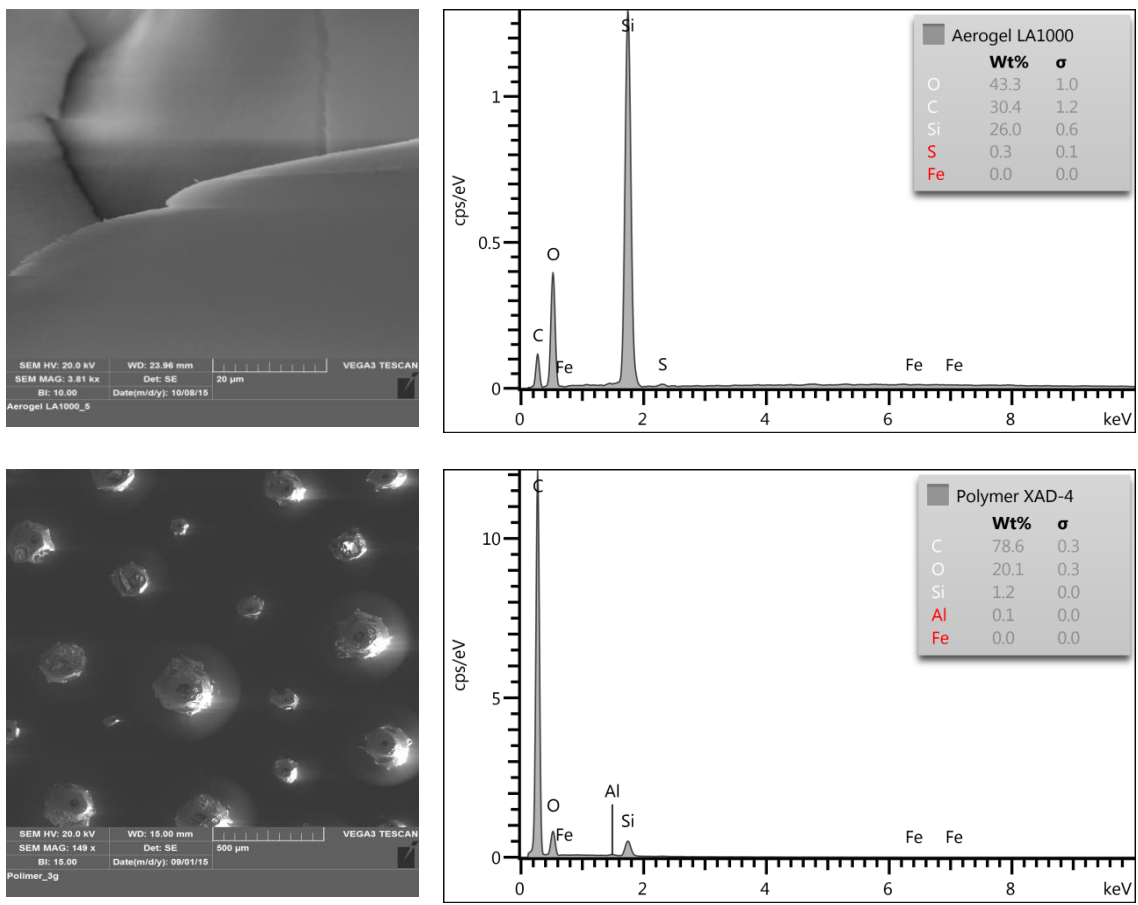

Fig.2 SEM images and EDS spectra acquired for aerogel LA-1000 (top row) and polymer XAD-4 (bottom).

Acknowledgments
We acknowledge technical and administrative support of A. Heczko, M. Kajetanowicz, G. Konopka-Cupiał, W. Migdał, and the financial support by the Polish National Centre for Research and Development through grant INNOTECH-K1/IN1/64/159174/NCBR/12, and grant LIDER/274/L-6/14/NCBR/2015, the Foundation for Polish Science through MPD programme, the EU and MSHE Grant No. POIG.02.03.00-161 00- 013/09, and The Marian Smoluchowski Krakow Research Consortium Matter-Energy-Future.